# Persistent gapless surface states in MnBi$_2$Te$_4$/Bi$_2$Te$_3$ superlattice antiferromagnetic topological insulator

L. X. Xu[1,2,3,4*], Y. H. Mao[5*], H. Y. Wang[2,4*], J. H. Li[6], Y. J. Chen[6], Y. Y. Y. Xia[2,4], Y. W. Li[7], J. Zhang[2], H. J. Zheng[2], K. Huang[2], C. F. Zhang[5], S. T. Cui[2], A. J. Liang[2,8], W. Xia[2,4], H. Su[2], S. W. Jung[9], C. Cacho[9], M. X. Wang[2], G. Li[2], Y. Xu[6,10], Y. F. Guo[2], L. X. Yang[6,10†], Z. K. Liu[2,3†], and Y. L. Chen[2,3,6,7†]

[1]*Center for Excellence in Superconducting Electronics, State Key Laboratory of Functional Material for Informatics, Shanghai Institute of Microsystem and Information Technology, Chinese Academy of Sciences, Shanghai 200050, China.*
[2]*School of Physical Science and Technology, ShanghaiTech University and CAS-Shanghai Science Research Center, Shanghai 201210, China.*
[3] *ShanghaiTech Laboratory for Topological Physics, Shanghai 200031, China.*
[4]*University of Chinese Academy of Sciences, Beijing 100049, China.*
[5]*College of Advanced Interdisciplinary Studies, National University of Defense Technology, Changsha 410073, China.*
[6]*State Key Laboratory of Low Dimensional Quantum Physics, Department of Physics, Tsinghua University, Beijing 100084, China.*
[7]*Department of Physics, Clarendon Laboratory, University of Oxford Parks Road, Oxford OX1 3PU, UK.*
[8]*Advanced Light Source, Lawrence Berkeley National Laboratory, Berkeley, California 94720, USA.*
[9]*Diamond Light Source, Harwell Campus, Didcot OX11 0DE, UK.*
10*Frontier Science Center for Quantum Information, Beijing 100084, China.*

*These authors contributed equally to this work.
†Email address: liuzhk@shanghaitech.edu.cn, lxyang@tsinghua.edu.cn, yulin.chen@physics.ox.ac.uk

**Magnetic topological quantum materials (TQMs) provide a fertile ground for the emergence of fascinating topological magneto-electric effects. Recently, the discovery of intrinsic antiferromagnetic (AFM) topological insulator MnBi$_2$Te$_4$ that could realize quantized anomalous Hall effect and axion insulator phase ignited intensive study on this family of TQM compounds. Here, we investigated the AFM compound MnBi$_4$Te$_7$ where Bi$_2$Te$_3$ and MnBi$_2$Te$_4$ layers alternate to form a superlattice. Using spatial- and angle-resolved photoemission spectroscopy, we identified ubiquitous (albeit termination dependent) topological electronic structures from both Bi$_2$Te$_3$ and MnBi$_2$Te$_4$ terminations. Unexpectedly, while the bulk bands show strong temperature dependence correlated with the AFM transition, the topological surface states show little temperature dependence and remain gapless across the AFM transition. The detailed electronic structure of MnBi$_4$Te$_7$ and its temperature evolution, together with the results of its sister compound MnBi$_2$Te$_4$, will not only help understand the exotic properties of this family of magnetic TQMs, but also guide the design for possible applications.**

# Introduction

Magnetism breaks the time reversal symmetry (TRS) that is crucial in the topological quantum materials (TQMs). Its interplay with topological electronic structures can foster profoundly interesting phenomena (e.g. quantum anomalous Hall effect (QAH), topological electromagnetic dynamics) and give rise to new states (e.g. magnetic Dirac and Weyl semimetals, axion insulators and topological superconductors [1-8]). Until recently, the introduction of magnetism into TQMs was typically achieved by magnetic doping in existing TQMs, such as the observation of QAH effect in V- or Cr-doped $(Bi_{1-x}Sb_x)_2Te_3$ thin films [9-11]. On the other hand, although there have been numerous intrinsic magnetic TQMs proposed theoretically [6, 12-18], their experimental realization are rare other than a few magnetic Weyl compounds [19-21].

Recently, a new magnetic topological insulator was theoretically proposed and soon experimentally verified in a layered compound $MnBi_2Te_4$, which possesses intrinsic antiferromagnetic (AFM) order[22-25] and provides a practical platform to realize many interesting phenomena and phases, such as quantized anomalous Hall effect [26], axion insulator phase[27] and high number Chern insulator phase[28] in its thin films. However, despite these exciting developments, there has been a fundamental puzzle unsolved regarding the topological electronic structure in $MnBi_2Te_4$, that its topological surface states (TSSs) were observed to show a diminished gap even at the AFM state [29-32], in contrast to the theoretical calculations that expect sizeable gaps [23]. This discrepancy suggests that the magnetic, electronic structures and their interplay in $MnBi_2Te_4$, particularly on surface, requires further investigations.

Interestingly, due to the van der Waals interaction between MnBi$_2$Te$_4$ (MBT) layers, it is possible to intercalate other thin films, such as Bi$_2$Te$_3$ (BT) layers to form a series of superlattice [MnBi$_2$Te$_4$]$_m$[Bi$_2$Te$_3$]$_n$ (m=1,2,3…, n=0,1,2,…). By controlling the ratio of the MnBi$_2$Te$_4$ and Bi$_2$Te$_3$ layer numbers, one can systematically tune the coupling between adjacent MnBi$_2$Te$_4$ layers and study the evolution of their magnetic properties and electronic structures. This flexibility makes this series of compounds an ideal platform for investigating the interplay between magnetism, topology and electronic structure in intrinsic magnetic topological insulators.

In this work, we focus our study in the MnBi$_4$Te$_7$ crystal, or the MnBi$_2$Te$_4$ : Bi$_2$Te$_3$=1:1 superlattice (see Fig. 1A). The fact that the two adjacent MnBi$_2$Te$_4$ layers are separated by a Bi$_2$Te$_3$ layer suggests a weakened magnetic coupling between MnBi$_2$Te$_4$ layers, with a lowered magnetic ordering temperature. Indeed, the AFM transition of MnBi$_4$Te$_7$ happens at $T_N$ ~ 12.5 K [33-37], which is about half of that ($T_N$ ~ 25 K) in MnBi$_2$Te$_4$ single crystals [25-27, 29, 31]. On the other hand, the weakened AFM interaction could also lead to easier introduction of the FM order, as suggested by a recent study that demonstrates an FM like magnetic transition at $T$c ~ 5K [34-36], making MnBi$_4$Te$_7$ a promising candidate for the realization of high temperature QAH effect.

Despite such importance and the enormous research efforts recently [33-37], The electronic structures of MnBi$_4$Te$_7$ remain controversial [34-36] and even puzzling (such as a gap in the TSSs as large as 90 meV was recently reported [34] at 300 K, which is much larger than the AFM ordering temperature of 12.5 K ). These controversies, as well as the intriguing differences between the electronic structures of MnBi$_2$Te$_4$ (reported to harbour gapless TSSs [29-32]) and MnBi$_4$Te$_7$ (reported to harbour gapped TSSs [34-36]), calls for a systematic investigation on the bulk and surface electronic structure of MnBi$_4$Te$_7$.

By combining synchrotron and laser light sources, we carried out comprehensive angle-resolved photoemission spectroscopy (ARPES) studies on MnBi$_4$Te$_7$, and identified its topological electronic structures including the characteristic TSSs. Furthermore, the high spatial resolution (~ 1 μm) in our experiments allowed us to identify the two terminations (MBT-terminated and BT-terminated) on the cleaved sample surfaces, both showing gapless TSSs within our energy resolution, similar to the gapless TSSs in MnBi$_2$Te$_4$ [31]. Finally, we carried out temperature-dependent measurements of both bulk and TSS bands, which shows intriguing difference across the magnetic phase transition: while the bulk states show a clear reduction of band splitting, the TSSs remain gapless. This difference indicates a complex influence of magnetism and interlayer coupling on the topological electronic structure in this family of magnetic topological insulators.

## Results

**Sample Characterization**

MnBi$_4$Te$_7$ crystallizes into a hexagonal lattice with space group of $P\bar{3}m1$ (No. 164). It is built up from stacking the van der Waals septuple Te-Bi-Te-Mn-Te-Bi-Te (MBT) and quintuple Te-Bi-Te-Bi-Te (BT) layers alternatively, as shown in Fig. 1A. The high quality of the single-crystalline samples used in this work is demonstrated by the single crystal X-ray diffraction patterns (Fig. 1B) and angle scan (Fig. 1C), showing no additional impurity phases other than MnBi$_4$Te$_7$. The magnetic susceptibility and electric transport measurements (Fig. 1D) are consistent with previous reports [34-36] and clearly show the

AFM phase transition at $T_N$ = 12.1 K. At low temperatures (< 5 K) and under magnetic field parallel to the $c$ axis, MnBi$_4$Te$_7$ undergoes a first-order spin-flip transition into a FM-like phase as manifested by the divergence between zero-field-cooling and field-cooling measurements (Fig. 1D). The $M$ - $H$ hysteresis saturates at $H_c$ = 0.22 T (Fig. 1E), which is about 40 times smaller than in MnBi$_2$Te$_4$. The relatively small saturation field suggests weaker interlayer magnetic coupling in MnBi$_4$Te$_7$ comparing to MnBi$_2$Te$_4$ and thus makes it a promising material candidate for QAH insulators under weak magnetic field.

Due to their two-dimensional nature, the samples could be cleaved between the MBT and BT layers, therefore leaving either MBT- or BT-terminated surfaces (Fig. 1A). As demonstrated by the atomic force microscope measurement, after cleaving, we have observed terrace heights that are consistent with the thicknesses of MBT, BT and MBT + BT layers (Figs. 1F and 1G). Therefore, it is possible to measure the electronic structures of both types of terminations on the same sample after a single cleavage.

**Electronic structures on different terminations**

Figure 2 presents the results of synchrotron-based spatial-resolved ARPES measurements (with beam spot size ~ 800 nm, the details of the measurement could be found in the supplemental materials). Two types of terminations [(1) and (2)] are well distinguished from the real-space intensity map (Fig. 2A right) showing clear contrast on Mn 3$d$ orbitals. In addition, clear chemical shifts (see Fig. 2B) are observed on Bi 5$d$ and Te 4$p$ peaks in the core level spectra from position (1) and position (2). According to the intensity of Mn 3$d$ orbital, we ascribe Mn-deficient area (1) and Mn-rich area (2) to BT- and MBT-terminated surfaces, respectively.

Figure 2C shows the typical electronic structures measured on BT- and MBT-terminated sample surfaces. The band dispersions in a large energy range (~ 5 eV) obtained from areas (1) and (2) share overall similarities [Figs. 2C(i) and (ii)]. Nevertheless, there are notable differences between them regarding the fine structure and relative intensity of the band dispersions. The most prominent distinction is the energy position and relative intensity of Mn 3$d$ bands which lie 3 ~ 4 eV below Fermi energy ($E_F$), proving areas (1) and (2) are BT and MBT terminations, respectively. The measured electronic structure contains both the electronic states from the bulk and surface. Fig. 2D shows the intensity map of the stacked momentum distribution curves (MDCs) at $E_F$ from the BT termination along the $\bar{M} - \bar{\Gamma} - \bar{M}$ direction using different photon energies, from which we could find that while the features located at $\pm\, 0.15$ Å$^{-1}$ show no clear variation with photon energy, the feature located around $\bar{\Gamma}$ shows alternation in the shape and intensity with photon energy, indicating their surface and bulk nature. The observed surface and bulk bands are identified as TSSs and bulk conduction band (CB), respectively (see below).

**Topological electronic structure**

The detailed electronic structures of MnBi$_4$Te$_7$ are investigated by high-resolution laser-based spatial-resolved (beam size ~ 20 μm) ARPES measurements ($hv = 6.994\ eV$) on both BT (see Figs. 3A-C) and MBT terminations (see Figs. 3E-G). The superb energy and momentum resolutions of laser-ARPES enable the revelation of petal-like hexagonal structures on the Fermi surface (FS) from both terminations as shown in Figs. 3A and 3E. The signature of the MBT-terminated FS is a large circular electron pocket (see Fig. 3E), which is absent on the BT termination. The constant energy contours on both terminations show similar conic evolution with binding energy (Figs. 3B and 3F). Figs. 3C and 3D (3G and 3H) compare the measured and calculated band dispersions of AFM MnBi$_4$Te$_7$ along $\bar{M} - \bar{\Gamma} - \bar{M}$ on BT (MBT) terminations. From the comparison, we can separate the bulk bands [CB1 - CB3 and valance bands (VB1 - VB2)] that do not show much variation with surface terminations from the termination-dependent surface bands (TSS$_a$ on BT termination and TSS$_b$ & SS$_b$ on MBT termination). We note that bulk bands also exhibit weak termination dependence by cautious analysis (especially the VB1 - VB2), which could be due to different projection weights and/or the variation of surface potential due to charge transfer on different terminations. The band inversion between the CB and VB are contributed by Te-$p_z$ and Bi-$p_z$ orbitals with opposite parities, similar to Bi$_2$Te$_3$[38], hence has a nontrivial topology. The measured bulk gap size is estimated to be ~ 100 meV, which is comparable to the value in MnBi$_2$Te$_4$.

On the BT termination, within our energy resolution, the TSS$_a$ is found to form a gapless structure in the bulk gap. The detailed analysis of the TSS$_a$ is plotted in Fig. 3C(iii), where the zoomed-in intensity plot (Fig. 3C (iii)) shows the linear dispersion around the Dirac point. The MDC (red curve) shows considerable intensity near the Dirac point and the EDC (green) across the Dirac point shows the single peak structure corresponding to the TSS$_a$. The EDC peak near the Dirac point can be fitted with a single Lorentzian peak as shown in the inset of Fig. 3C(iii). The detailed analysis supports a gapless structure of TSS$_a$ within the energy resolution of our laser-ARPES (~ 3 meV) near the Dirac point sitting at 0.275 eV below $E_F$.

The TSSs can be reproduced in the slab calculation as shown in Fig. 3D. Notably, the lower branch of the TSSs is merged into the VB in the calculation (Fig. 3D). By including an onsite electrostatic voltage that is possibly induced by electron transfer process on the BT termination, the TSSs are well separated from the bulk bands and a gapless TSS presents (See supplemental materials). Such gapless TSSs on the surface of AFM MnBi$_4$Te$_7$ could be explained by the attenuation of the TRS breaking magnetization across the top BT layer that is insufficiently strong for gap opening in the TSSs. Interestingly, we observe a substantial band hybridization between bulk CB2 and surface TSS$_a$, which suggests the petal-like hexagonal feature on the FS has the contribution from both TSS$_a$ and CB2.

Similarly, we observe gapless TSSs on the MBT termination. The detailed analysis of the TSS$_b$ are presented in Fig. 3G(iii), including the zoomed-in intensity plot, the MDC around the Dirac point as well as the EDC at $\bar{\Gamma}$ that can be well fitted by a single Lorentzian peak near the Dirac point, all supporting a gapless structure in the TSSs with the Dirac point sitting at 0.275eV below $E_F$, similar to the TSS$_a$ on the BT termination. Such observation, however, is beyond the expectation of *ab initio* calculation that predicts TSSs with an energy gap of about 28 meV and inconsistent with previous synchrotron-based ARPES measurements [36]. Such deviation might be resulted from similar mechanisms of gapless TSS observed in the sister compound MnBi$_2$Te$_4$, where the extended wave function of TSSs is mediated by the average effect of different AFM domains or the disordered surface magnetization [13, 29, 31]. Besides, considering the much smaller energy differences between the magnetic configurations in MnBi$_4$Te$_7$ [36], local impurities and fluctuation could easily lead to nano-sized domains of magnetic order other than the theoretically proposed *A*-AFM configuration [39]. Therefore, the TSSs remain gapless in magnetic domains that respect the TRS (see discussion and supplemental materials for details).

**Temperature evolution of the electronic structure of MnBi$_4$Te$_7$**

To investigate the response of topological electronic structure in different magnetic phases, we carried out systematic temperature-dependent ARPES measurements across $T_N$, which are depicted in Fig. 4 (also see supplemental materials). Across the $T_N$, we observed a dual temperature evolution of the bulk states and TSSs (Fig. 4A). As for the TSSs, we note the shape of the dispersion remains from the zoomed-in intensity plots around the Dirac point (Fig. 4B) and the EDC at the $\bar{\Gamma}$ point keeps the single-peak structure at the Dirac point (Fig. 4C), all suggesting the gapless nature of TSSs remains across the antiferromagnetic phase transition. The unexpected robustness of the topological electronic structure of MnBi$_4$Te$_7$ regardless of the magnetic ordering mimics the situation in MnBi$_2$Te$_4$.

In contrast, we have found clear band evolution in both CB and VB. The difference of the band structure is highlighted by the comparison of the second derivative plot of photoemission intensity measured at 7.5 K (AFM phase) and 13.5 K (close to transition) and 17 K (PM phase) (Fig. 4D). Evidently, across the AFM-PM phase transition, the VB1 and VB2 bands shift towards each other while the hybridization between CB2 and TSS$_a$ bands diminishes due to the relative shift between CB2 and TSS$_a$, as illustrated Fig. 4D and supplemental materials. We track the temperature evolution of the above-mentioned bands with their EDC peaks at various momenta and different temperatures, as shown in Fig. 4E (and Fig. S4). From the extracted peak positions, we found the splitting between VB1 and VB2 reduces to minimum above the $T_N$, as summarized in Figs. 4E and 4F. The band changes across the AFM transition suggest the keen correlation between the electronic structure and magnetic ordering in MnBi$_4$Te$_7$. Especially, the reduction of the splitting between VB1 and VB2 maybe due to exchange splitting from the FM MBT layer [31, 40] or the disappearing of the $k_z$ folding effect across the AFM-PM phase transition (see Fig. S7-2 in supplemental materials).

## Discussion

In contrast to previous first principle calculation and synchrotron-based ARPES studies [34-36], our laser-ARPES with superb energy and momentum resolutions identified gapless TSSs on both BT- and MBT-terminated surfaces. While the gapless TSS on the BT termination is preserved due to intact TRS considering the attenuation of the AFM magnetisation across the non-magnetic Bi$_2$Te$_3$ layer (as reproduced by our *ab initial* calculation in Fig. 3D and Fig. S6), the gapless TSS on the magnetic MBT termination is beyond the expectation of the calculation that suggests a gapped TSS with an energy gap of about 28 meV caused by the broken TRS (Fig. 3H). This observation resembles the behavior in MnBi$_2$Te$_4$ system, in which a gapless TSS is attributed to the disordering of magnetisation or the mediation of the delocalized TSS by multi-AFM/FM-domains with anisotropic orientations of magnetisation [31]. Since the energy difference among distinct AFM/FM configurations are even much smaller in MnBi$_4$Te$_7$ than in MnBi$_2$Te$_4$ (e.g., for example, the FM phase has energy of ~ 0.5 meV/Mn higher than the lowest *A*-AFM phase, 10 times smaller than that in MnBi$_2$Te$_4$; the energy difference between *A*-type and *X*-type AFM is about 1.7 meV in MnBi$_2$Te$_4$ [23] and 0.5 meV in MnBi$_4$Te$_7$ [36]), multiple energetically favourable AFM phases with negligible energy difference are likely to coexist in MnBi$_4$Te$_7$, contributing to the gapless TSS on the MBT termination.

On the other hand, the identification of termination-dependent TSSs provides a platform to further explore rich and fascinating topological phenomena. Unlike the situation in MnBi$_2$Te$_4$ that shows a single type of termination, heterostructures with different terminations can be easily fabricated by exfoliating MnBi$_4$Te$_7$ into ultrathin films. Due to their distinctive topological electronic structure, interesting topological magnetoelectric effects can be expected [34, 35, 41]. In particular, the additional Bi$_2$Te$_3$ layer inserted between MnBi$_4$Te$_7$ layers strongly suppresses the interlayer AFM coupling between MnBi$_2$Te$_4$ layers. Therefore, it will be much easier to align the magnetisation in MnBi$_4$Te$_7$ to form FM magnetic phase and realize low- and even zero-field QAH effect. Actually, magnetization measurement has revealed a FM-like ground state below 5 K (Fig. 1E) [34, 35], in which either a magnetic Weyl semimetal or topological crystalline insulator phase has been predicted by *ab initio* calculation. The complex and intriguing magnetic ordering in MnBi$_4$Te$_7$ will provide a rich platform to investigate the interplay between magnetism and topological electronic structure and search for more topological magnetoelectric phenomena (e.g, topological phase transitions).

Finally, our temperature-dependent measurement shows delicate electronic structure evolution across AFM/PM transition in MnBi$_4$Te$_7$. Interestingly, while the bulk bands show strong temperature dependence, the surface states on both terminations remain gapless across AFM transition and the predicted alternation of the dispersion of the TSSs is absent in the experiment [see Fig. 4 and the supplemental materials for details], resembling the dual behavior of surface and bulk bands in MnBi$_2$Te$_4$ [31]. However, the rich termination-dependent surface states provide a fecund playground to investigate the crucial role of the AFM ordering on the TSSs in MnBi$_4$Te$_7$. As observed in Fig. 4, the surface $TSS_a$ hybridizes with bulk CB2 in the AFM phase but is decoupled above $T_N$. The dispersion of CB2 is influenced by AFM ordering via introducing overlapping with $TSS_a$ and thus realizes indirect modulation of the TSS near $E_F$.

## Conclusion

In conclusion, $MnBi_2Te_4/Bi_2Te_3$ superlattice is realized in stoichiometric single crystal $MnBi_4Te_7$. We observe termination-dependent, yet robust gapless TSSs, which identifies the topological nature of this novel superlattice compound. Interestingly, we observe strong modulation of the bulk bands by the AFM ordering below $T_N$, while the TSSs unexpectedly show negligible temperature dependence and remain gapless across $T_N$. Our results will not only help understand the novel properties of the $[MnBi_2Te_4]_m[Bi_2Te_3]_n$ superlattice compounds, but also guide the exploration of rich topological physics in $MnBi_2Te_4/Bi_2Te_3$ heterostructures.

## Materials and Methods

### Crystal growth

$MnBi_4Te_7$ single crystals were grown by using the self-flux method. Mn (99.95%) and $Bi_2Te_3$ (99.999%) pieces were mixed together in a molar ratio of Mn : $Bi_2Te_3$ = 1:6. The mixture was placed into an alumina crucible and sealed in a silica tube under vacuum. The tube put into a furnace was slowly heated up to 700 °C and kept at this temperature for more than 30 hours. The assembly was subsequently slowly cooled to 590 °C at a rate of 5 °C/h, and then slowly cooled to 580 °C at a rate of 1 °C/h. It was finally taken out of the furnace at 580 °C and was put into a high-speed centrifuge immediately to remove the excess flux. Clean MnBi$_4$Te$_7$ single crystals were thus left in the aluminium crucible.

**Angle-resolved photoemission spectroscopy**

ARPES measurements were performed at both high-resolution and Nano branches of beamline I05 (proposal nos. SI23648 and SI24827) in the Diamond Light Source (DLS), UK. The samples were cleaved *in situ* and measured under ultra-high vacuum below $1 \times 10^{-10}$ Torr at DLS. Data were collected by Scienta R4000 and DA30L analysers at HR and Nano branch, respectively. The total energy and angle resolutions were 15 meV and 0.2 °, respectively.

High-resolution laser-based ARPES measurements were performed at home-built setups ($h\nu = 6.994\ eV$) in ShanghaiTech and Tsinghua University. The samples were cleaved *in situ* and measured under ultra-high vacuum below $6 \times 10^{-11}$ Torr. Data were collected by a DA30 analyser. The total energy and angle resolutions were 2.5 meV and 0.2 °, respectively.

**First-principles calculations**

First-principles calculations were performed by density functional theory (DFT) using the Vienna *ab-initio* Simulation Package. The plane-wave basis with an energy cutoff of 350 eV was adopted. The electron-ion interactions were modeled by the projector augmented wave potential and the exchange-correlation functional was approximated by the Perdew-Burke-Ernzerhof type generalized gradient approximation (GGA) [42]. The GGA+U method was applied to describe the localized 3$d$-orbitals of Mn atoms, for which U = 4.0 eV was selected according to our previous tests [22]. The structural relaxation for optimized lattice constants and atomic positions was performed with a force criterion of

0.01 eV/ Å and by using the DFT-D3 method to include van der Waals corrections. Spin-orbit coupling was included in self-consistent calculations and the Monkhorst-Pack k-point mesh of $9 \times 9 \times 3$ was adopted. Surface state calculations were performed with WannierTools package [43], based on the tight-binding Hamiltonians constructed from maximally localized Wannier functions (MLWF) [44].

## References and Notes

[1] J.E. Moore, Nature **464**, 194-198 (2010).

[2] R. Li, J. Wang, X.-L. Qi, S.-C. Zhang, Nature Physics **6**, 284-288 (2010).

[3] M.Z. Hasan, C.L. Kane, Reviews of Modern Physics **82**, 3045-3067 (2010).

[4] Y.L. Chen, J.-H. Chu, J.G. Analytis, Z.K. Liu, K. Igarashi, H.-H. Kuo, X.L. Qi, S.K. Mo, R.G. Moore, D.H. Lu, M. Hashimoto, T. Sasagawa, S.C. Zhang, I.R. Fisher, Z. Hussain, Z.X. Shen, Science **329**, 659 (2010).

[5] X.-L. Qi, T.L. Hughes, S.-C. Zhang, Physical Review B **82**, (2010).

[6] R. Yu, W. Zhang, H. Zhang, S.-c. Zhang, X. Dai, Z. Fang, Science **329**, 61 (2010).

[7] X.-L. Qi, S.-C. Zhang, Physics Today **63**, 33-38 (2010).

[8] X.-L. Qi, S.-C. Zhang, Rev. Mod. Phys. **83**, 1057-1110 (2011).

[9] C.-Z. Chang, J. Zhang, X. Feng, J. Shen, Z. Zhang, M. Guo, K. Li, Y. Ou, P. Wei, L.-L. Wang, Z.-Q. Ji, Y. Feng, S. Ji, X. Chen, J. Jia, X. Dai, Z. Fang, S.-C. Zhang, K. He, Y. Wang, L. Lu, X.-C. Ma, Q.-K. Xue, Science **340**, 167-170 (2013).

[10] C.-Z. Chang, W. Zhao, D.Y. Kim, H. Zhang, B.A. Assaf, D. Heiman, S.-C. Zhang, C. Liu, M.H.W. Chan, J.S. Moodera, Nat. Mater. **14**, 473 (2015).

[11] X. Kou, S.-T. Guo, Y. Fan, L. Pan, M. Lang, Y. Jiang, Q. Shao, T. Nie, K. Murata, J. Tang, Y. Wang, L. He, T.-K. Lee, W.-L. Lee, K.L. Wang, Physical Review Letters **113**, 137201 (2014).

[12] P. Tang, Q. Zhou, G. Xu, S.-C. Zhang, Nat. Phys. **12**, 1100 (2016).

[13] J. Wang, arXiv:1701.00896 (2017).

[14] X. Wan, A.M. Turner, A. Vishwanath, S.Y. Savrasov, Phys. Rev. B **83**, 205101 (2011).

[15] G. Xu, H. Weng, Z. Wang, X. Dai, Z. Fang, Phys. Rev. Lett. **107**, 186806 (2011).

[16] E. Liu, Y. Sun, N. Kumar, L. Muechler, A. Sun, L. Jiao, S.-Y. Yang, D. Liu, A. Liang, Q. Xu, J. Kroder, V. Süß, H. Borrmann, C. Shekhar, Z. Wang, C. Xi, W. Wang, W. Schnelle, S. Wirth, Y. Chen, S.T.B. Goennenwein, C. Felser, Nat. Phys. **14**, 1125-1131 (2018).

[17] Q. Wang, Y. Xu, R. Lou, Z. Liu, M. Li, Y. Huang, D. Shen, H. Weng, S. Wang, H. Lei, Nat. Commun. **9**, 3681 (2018).

[18] R.S.K. Mong, A.M. Essin, J.E. Moore, Phys. Rev. B **81**, 245209 (2010).

[19] I. Belopolski, K. Manna, D.S. Sanchez, G. Chang, B. Ernst, J. Yin, S.S. Zhang, T. Cochran, N. Shumiya, H. Zheng, B. Singh, G. Bian, D. Multer, M. Litskevich, X. Zhou, S.-M. Huang, B. Wang, T.-R. Chang, S.-Y. Xu, A. Bansil, C. Felser, H. Lin, M.Z. Hasan, Science **365**, 1278-1281 (2019).

[20] D.F. Liu, A.J. Liang, E.K. Liu, Q.N. Xu, Y.W. Li, C. Chen, D. Pei, W.J. Shi, S.K. Mo, P. Dudin, T. Kim, C. Cacho, G. Li, Y. Sun, L.X. Yang, Z.K. Liu, S.S.P. Parkin, C. Felser, Y.L. Chen, Science **365**, 1282-1285 (2019).

[21] N. Morali, R. Batabyal, P.K. Nag, E. Liu, Q. Xu, Y. Sun, B. Yan, C. Felser, N. Avraham, H. Beidenkopf, Science **365**, 1286-1291 (2019).

[22] J. Li, Y. Li, S. Du, Z. Wang, B.-L. Gu, S.-C. Zhang, K. He, W. Duan, Y. Xu, Science Advances **5**, eaaw5685 (2019).

[23] D. Zhang, M. Shi, T. Zhu, D. Xing, H. Zhang, J. Wang, Physical Review Letters **122**, 206401 (2019).

[24] M.M. Otrokov, I.P. Rusinov, M. Blanco-Rey, M. Hoffmann, A.Y. Vyazovskaya, S.V. Eremeev, A. Ernst, P.M. Echenique, A. Arnau, E.V. Chulkov, Physical Review Letters **122**, 107202 (2019).

[25] J.Q. Yan, Q. Zhang, T. Heitmann, Z. Huang, K.Y. Chen, J.G. Cheng, W. Wu, D. Vaknin, B.C. Sales, R.J. McQueeney, Physical Review Materials **3**, (2019).

[26] Y. Deng, Y. Yu, M.Z. Shi, J. Wang, X.H. Chen, Y. Zhang, arXiv:1904.11468 (2019).

[27] C. Liu, Y. Wang, H. Li, Y. Wu, Y. Li, J. Li, K. He, Y. Xu, J. Zhang, Y. Wang, arXiv:1905.00715 (2019).

[28] J. Ge, Y. Liu, J. Li, H. Li, T. Luo, Y. Wu, Y. Xu, J. Wang, arXiv:1907.09947 (2019).

[29] Y.-J. Hao, P. Liu, Y. Feng, X.-M. Ma, E.F. Schwier, M. Arita, S. Kumar, C. Hu, R.e. Lu, M. Zeng, Y. Wang, Z. Hao, H. Sun, K. Zhang, J. Mei, N. Ni, L. Wu, K. Shimada, C. Chen, Q. Liu, C. Liu, arXiv:1907.03722 (2019).

[30] P. Swatek, Y. Wu, L.-L. Wang, K. Lee, B. Schrunk, J. Yan, A. Kaminski, arXiv:1907.09596 (2019).

[31] Y.J. Chen, L.X. Xu, J.H. Li, Y.W. Li, C.F. Zhang, H. Li, Y. Wu, A.J. Liang, C. Chen, S.W. Jung, C. Cacho, H.Y. Wang, Y.H. Mao, S. Liu, M.X. Wang, Y.F. Guo, Y. Xu, Z.K. Liu, L.X. Yang, Y.L. Chen, arXiv:1907.05119 (2019).

[32] H. Li, S.-Y. Gao, S.-F. Duan, Y.-F. Xu, K.-J. Zhu, S.-J. Tian, W.-H. Fan, Z.-C. Rao, J.-R. Huang, J.-J. Li, Z.-T. Liu, W.-L. Liu, Y.-B. Huang, Y.-L. Li, Y. Liu, G.-B. Zhang, H.-C. Lei, Y.-G. Shi, W.-T. Zhang, H.-M. Weng, T. Qian, H. Ding, arXiv:1907.06491 (2019).

[33] L. Ding, C. Hu, F. Ye, E. Feng, N. Ni, H. Cao, Crystal and magnetic structure of magnetic topological insulators MnBi2nTe3n+1. arXiv:1910.06248, (2019)

[34] J. Wu, F. Liu, M. Sasase, K. Ienaga, Y. Obata, R. Yukawa, K. Horiba, H. Kumigashira, S. Okuma, T. Inoshita, H. Hosono, arXiv:1905.02385 (2019).

[35] C. Hu, X. Zhou, P. Liu, J. Liu, P. Hao, E. Emmanouilidou, H. Sun, Y. Liu, H. Brawer, A.P. Ramirez, H. Cao, Q. Liu, D. Dessau, N. Ni, arXiv:1905.02154 (2019).

[36] R.C. Vidal, A. Zeugner, J.I. Facio, R. Ray, M.H. Haghighi, A.U.B. Wolter, L.T.C. Bohorquez, F. Caglieris, S. Moser, T. Figgemeier, T.R.F. Peixoto, H.B. Vasili, M. Valvidares, S.

Jung, C. Cacho, A. Alfonsov, K. Mehlawat, V. Kataev, C. Hess, M. Richter, B. Büchner, J.v.d. Brink, M. Ruck, F. Reinert, H. Bentmann, A. Isaeva, arXiv:1906.08394 (2019).

[37] J.-Q. Yan, Y. H. Liu, D. Parker, M. A. McGuire, B. C. Sales, A-type Antiferromagnetic order in $MnBi_4Te_7$ and $MnBi_6Te_{10}$ single crystals. arXiv:1910.06273 (2019)

[38] Y.L. Chen, J.G. Analytis, J.-H. Chu, Z.K. Liu, S.-K. Mo, X.L. Qi, H.J. Zhang, D.H. Lu, X. Dai, Z. Fang, S.C. Zhang, I.R. Fisher, Z. Hussain, Z.-X. Shen, science **325**, 5937 (2009).

[39] D.P. J.-Q. Yan, 1 Liqin Ke,2 A.-M. Nedić,3 Y. Sizyuk,2 Elijah Gordon,2 P. P. Orth,2, 3 D. Vaknin,2, 3 and R. J. McQ.

[40] M. M. Otrokov, I. I. Klimovskikh, H. Bentmann, A. Zeugner, Z. S. Aliev, S. Gass, A. U. B. Wolter, A. V. Koroleva, D. Estyunin, A. M. Shikin, M. Blanco-Rey, M. Hoffmann, A. Y. Vyazovskaya, S. V. Eremeev, Y. M. Koroteev, I. R. Amiraslanov, M. B. Babanly, N. T. Mamedov, N. A. Abdullayev, V. N. Zverev, B. Büchner, E. F. Schwier, S. Kumar, A. Kimura, L. Petaccia, G. D. Santo, R. C. Vidal, S. Schatz, K. Kißner, C.-H. Min, S. K. Moser, T. R. F. Peixoto, F. Reinert, A. Ernst, P. M. Echenique, A. Isaeva, E. V. Chulkov, Prediction and observation of the first antiferromagnetic topological insulator. arXiv:1809.07389, (2018).

[41] H. Sun, B. Xia, Z. Chen, Y. Zhang, P. Liu, Q. Yao, H. Tang, Y. Zhao, H. Xu, Q. Liu, Phys. Rev. Lett. **123**, 096401 (2019).

[42] J.P. Perdew, K. Burke, M. Ernzerhof, Phys. Rev. Lett. **77**, 3865-3868 (1996).

[43] Q. Wu, S. Zhang, H.-F. Song, M. Troyer, A.A. Soluyanov, Comput. Phys. Commun. **224**, 405-416 (2018).

[44] A.A. Mostofi, J.R. Yates, G. Pizzi, Y.-S. Lee, I. Souza, D. Vanderbilt, N. Marzari, Comput. Phys. Commun. **185**, 2309-2310 (2014).

## Acknowledgments

This work was supported by the National Key R&D program of China (Grant No. 2017YFA0305400 and 2017YFA0304600), the National Natural Science Foundation of China (Grant No. 11774190, No. 11674229, No. 11634009, and No. 11774427) and EPSRC Platform Grant (Grant No. EP/M020517/1). Y. F. G. acknowledges the support from the Shanghai Pujiang Program (Grant No. 17PJ1406200). L. X. Y. and Y. L. C. acknowledge the support from Tsinghua University Initiative Scientific Research Program. Y. W. L. acknowledges the support from China Scholarship Council.

**Figures and legends**

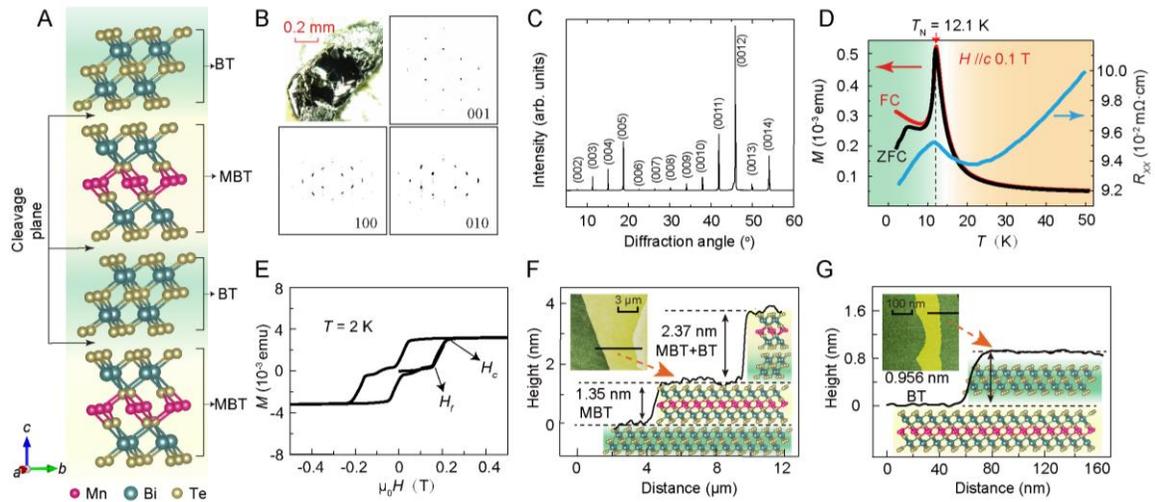

**Fig. 1. Basic properties and sample characterization of MnBi$_4$Te$_7$ single crystal.** (**A**) Schematic illustration of crystal structure (2 unit cells) of MnBi$_4$Te$_7$. Layered crystal structures, including the Te-Bi-Te-Bi-Te quintuple layer (BT), Te-Bi-Te-Mn-Te-Bi-Te septuple layer (MBT) as well as the cleavage planes are labelled. (**B**) Photos of MnBi$_4$Te$_7$ single crystal after cleavage and X-ray diffraction patterns along different directions. (**C**) Angle-scan of X-ray diffraction along *c*-axis. (**D**) Zero-field cooled (ZFC) and field-cooled (FC, with 0.1 T magnetic field applied along the *c* axis) magnetization together with resistivity (blue curve) as functions of temperature. The blue shade indicates an antiferromagnetic (AFM) phase while the orange shade indicates the paramagnetic (PM) phase. (**E**) Full magnetic hysteresis loop of isothermal magnetization taken at 2 K for *H* ∥ *c*. $H_f$ indicates the magnetic field when the first plateau is reached. $H_c$ indicates the saturated magnetic field. (**F**, **G**) AFM topography image (insets) with a respective section graphs of cleaved MnBi$_4$Te$_7$ surfaces, showing the characteristic BT, MBT and BT+MBT termination layers.

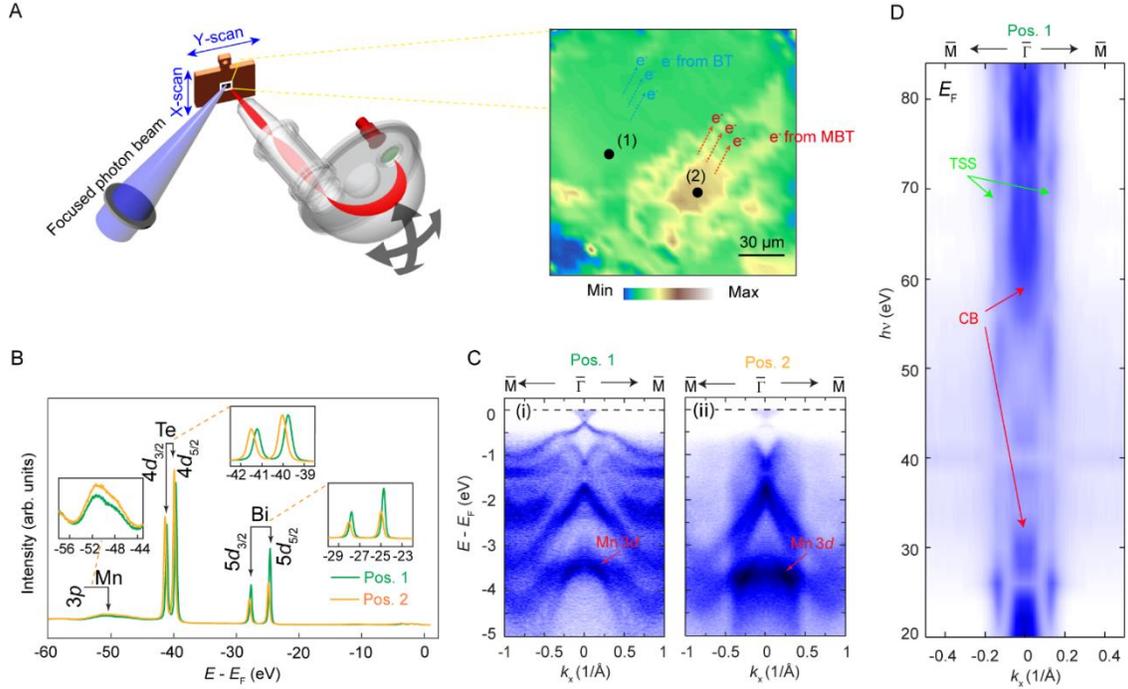

**Fig. 2. The electronic structure of the BT and MBT terminations in MnBi$_4$Te$_7$.** (**A**) Left: Schematic of the spatially resolved ARPES measurement in I05 beamline of DLS. Right: Sample surface mapping of ARPES intensity integrated in an energy window of ~ 1 eV centered at 4eV below $E_F$. Positions (1) and (2) indicate the Mn-deficient BT and Mn-rich MBT terminations, respectively. (**B**) The core-level spectrum of MnBi$_4$Te$_7$ at positions (1) and (2) with the characteristic element peaks labelled. Insets are the zoomed-in plots of the element peaks for comparison. (**C**) The overall photoemission intensity along the $\bar{M} - \bar{\Gamma} - \bar{M}$ direction measured at positions (1) (i) and (2) (ii), respectively. The position of the Mn 3$d$ band is labelled; Data were taken using 84 eV photons. (**D**) Intensity plot of stacked momentum distribution curves (MDCs) along the $\bar{M} - \bar{\Gamma} - \bar{M}$ direction at position (1) from photon-energy-dependent measurement under 20 to 86 eV near Fermi energy. The measurement temperature was kept at 35 K in (A)-(C) and 7 K in (D).

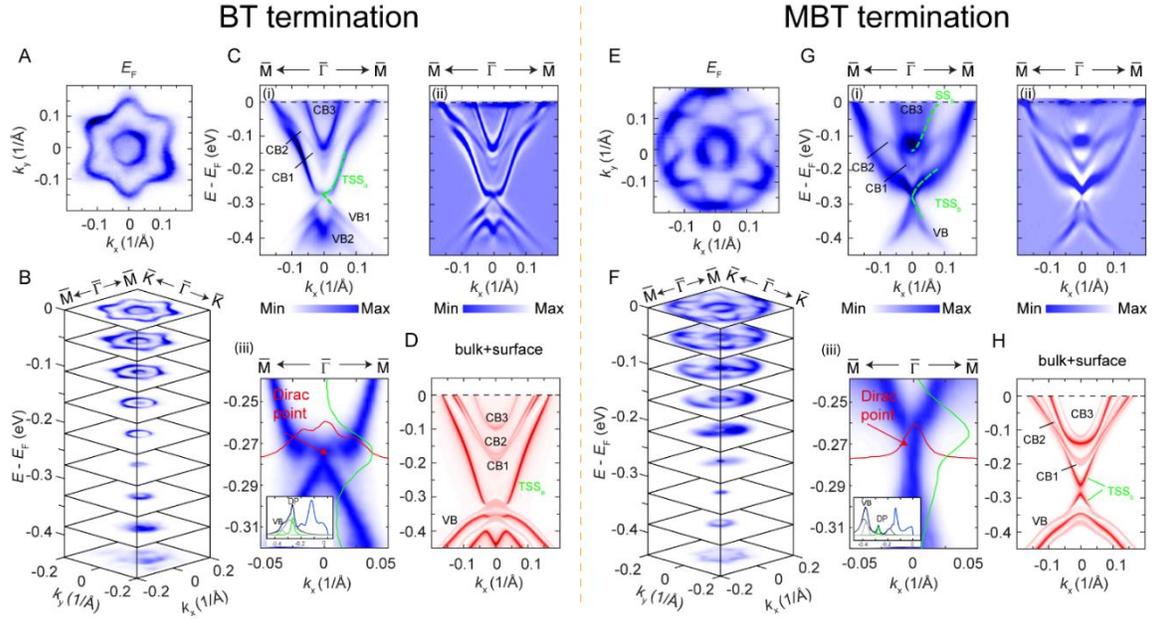

**Fig. 3. Band structure of MnBi$_4$Te$_7$ measured with 7 eV laser and comparison with calculation.** (**A, B**) Constant energy contour on the BT termination near the Fermi energy (A) and stacked plots at different binding energies (B). (**C**) Photoemission intensity on the BT termination along the $\bar{M} - \bar{\Gamma} - \bar{M}$ direction (i) and its second-derivative $\frac{d^2 I}{d\omega^2}$ plot (ii). Different dispersions were identified and labelled in (i). (iii) The zoom-in plot of the band dispersions near the Dirac point displaying gapless surface electronic structure. The red and green curve indicate the MDC and energy distribution curve (EDC) across the Dirac point, respectively. Inset: EDC across the Dirac point fitted with multi-Lorentzian peaks. A single EDC peak is observed near the Dirac point, confirming the gapless nature of the TSSs. (**D**) The *ab-initio* calculation of the projected bulk and surface electronic states along $\bar{M} - \bar{\Gamma} - \bar{M}$ direction. Different bands were labelled. (**E-H**) Same as (A-D), but on the MBT termination. The sample temperature was kept at 7.5 K for BT termination and 8.5 K for MBT termination.

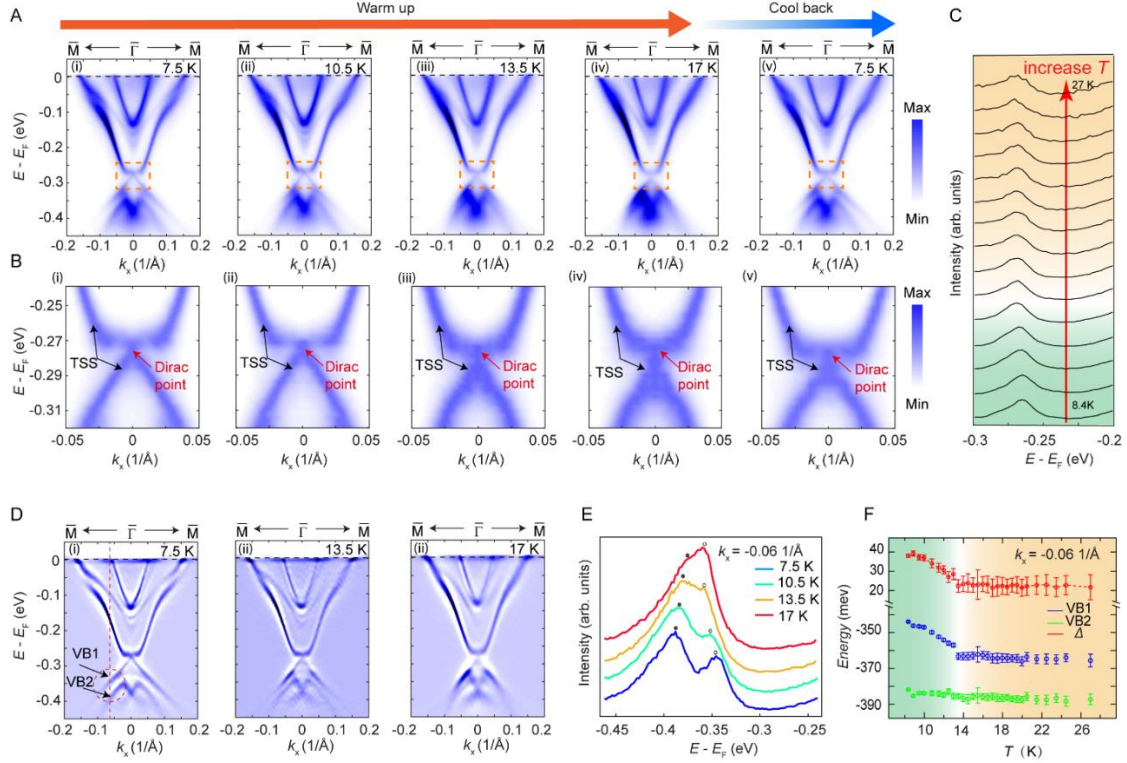

**Fig. 4. Temperature evolution of the band structure of BT-terminated MnBi$_4$Te$_7$.** (**A, B**) Band dispersions along the $\bar{M}-\bar{\Gamma}-\bar{M}$ direction (A) and corresponding zoom-in plots near the Dirac point (B) showing gapless surface electronic structure at selected temperatures. (**C**) Temperature evolution of EDCs at $\bar{\Gamma}$ around the Dirac point. (**D**) Side-by-side comparison of the second derivative $\frac{d^2 I}{d\omega^2}$ plot of photoemission intensity map along the $\bar{M}-\bar{\Gamma}-\bar{M}$ direction at 7.5 K (i), 13.5 K (ii), and 17 K (iii), respectively. Arrows indicate the VB1 and VB2 bands showing the most significant changes. Dashed lines indicate the positions of EDCs analyzed in (E, F). (**E**) Stacked plot of EDCs taken at $k_x = -0.06$ Å$^{-1}$ from (A). The solid and empty black circles indicate the fitted peak positions corresponding to the VB1 and VB2 bands. (**F**) The energy positions of VB1 (blue curve) and VB2 (green curve) bands together with the energy difference between them (red curve) as functions of the temperature. Data were taken with 7 eV laser.